\documentclass[a4paper,10pt]{article}
\usepackage{graphicx}
\usepackage{amssymb}

\title{Contour and surface integrals in potential scattering}
\date{5 February 2020}

\author{Giampiero Esposito, INFN Sezione di Napoli, \\
Complesso Universitario di Monte S. Angelo, \\
Via Cintia, Edificio 6, 80126 Napoli, Italy}

\begin{document}
\maketitle
\begin{abstract}
When the Schr\"{o}dinger equation for stationary states is studied for a system
described by a central potential in $n$-dimensional Euclidean space, the radial
part of stationary states is an even function of a parameter $\lambda$ which is
a linear combination of angular momentum quantum number $l$ and dimension $n$,
i.e., $\lambda=l+{(n-2)\over 2}$. Thus, without setting a priori $n=3$, complex
values of $\lambda$ can be achieved, in particular, by keeping $l$ real and
complexifying $n$. For suitable values of such an auxiliary complexified 
dimension, it is therefore possible to obtain results on scattering amplitude
and phase shift that are completely equivalent to the results obtained in
the sixties for Yukawian potentials in ${\bf R}^{3}$. Moreover, if both $l$ and 
$n$ are complexified, the possibility arises of recovering the partial wave
amplitude from residues of a function of two complex variables. Thus, the
complex angular momentum formalism can be imbedded into a broader framework, 
where a correspondence exists between the scattering amplitude and a skew
curve in ${\bf R}^{3}$.
\end{abstract}

\section{Introduction}
\setcounter{equation}{0}

The theory of potential scattering has been always of fundamental importance in
quantum mechanics, since it provides the appropriate tool for the investigation 
of matter at nuclear and subnuclear level. In this paper, we are interested in
obtaining a new perspective on potential scattering in ordinary 
(i.e., non-relativistic) quantum mechanics. For this purpose, our starting point 
is the familiar form of the scattering amplitude when the partial-wave expansion is 
applied in presence of Yukawian central potentials, i.e.
\cite{R1959,BLR1962,DR1965}
\begin{equation}
f(k,\cos \theta)={1 \over 2ik} \sum_{l=0}^{\infty}(2l+1)
\Bigr(e^{2i \delta(l,k)}-1 \Bigr) P_{l}(\cos \theta),
\label{(1.1)}
\end{equation}
where $\theta$ is the angle between initial wave vector ${\vec k}_{{\rm i}}$ 
and final wave vector ${\vec k}_{{\rm f}}$, 
$\delta(l,k)$ is the phase shift and $P_{l}$ is the standard notation 
for Legendre polynomials.
The work in Refs. \cite{R1959,BLR1962,DR1965} had the merit, among the many, of
recognizing that all terms in this sum can be obtained by applying the residue
theorem to a contour integral, where the function to be integrated is
\begin{equation}
Q_{k}: \lambda \in {\bf C} \rightarrow Q_{k}(\lambda)
={1 \over 2k} \Bigr(e^{2i \delta(\lambda,k)}-1 \Bigr)
{\lambda P_{\lambda - {1 \over 2}}(-\cos \theta)\over \cos (\pi \lambda)},
\label{(1.2)}
\end{equation}
and the integration path $\gamma$ encircles all zeros of 
$\cos (\pi \lambda)$ on the positive half-line 
${\rm Re}(\lambda)>0$, which are given by $l+{1 \over 2}$,
$\forall l=0,1,..., \infty$. Indeed, one finds
\begin{eqnarray}
\; & \; &
\left . {\rm Res} Q_{k}(\lambda) \right
|_{\lambda=l+{1 \over 2}}
=\lim_{\lambda \to l+{1 \over 2}}
\left(\lambda-l-{1 \over 2}\right)Q_{k}(\lambda)
\nonumber \\
&=& {(2l+1)\over 4k}\Bigr(e^{2i \delta(l,k)}-1 \Bigr)
P_{l}(-\cos \theta)
\lim_{\varepsilon \to 0}{\varepsilon \over
\cos \left(\pi l + {\pi \over 2} + \pi \varepsilon \right)}
\nonumber \\
&=& {(2l+1)\over 4k}\Bigr(e^{2 i \delta(l,k)}-1 \Bigr)
P_{l}(-\cos \theta) 
\lim_{\varepsilon \to 0}{\varepsilon \over
\Bigr[-\pi \varepsilon \cos(\pi l)\Bigr]}
\nonumber \\
&=&-{(2l+1)\over 4 \pi k}\Bigr(e^{2i \delta(l,k)}-1 \Bigr)
P_{l}(\cos \theta),
\label{(1.3)}
\end{eqnarray}
since $\cos(\pi l)=(-1)^{l}$, and 
$(-1)^{l}P_{l}(-\cos \theta)=P_{l}(\cos \theta)$. One 
therefore finds that
\begin{eqnarray}
\; & \; & \int_{\gamma}Q_{k}(\lambda)d\lambda =
2 \pi i \sum_{l=0}^{\infty} \left[
-{(2l+1)\over 4 \pi k} 
\Bigr(e^{2i \delta(l,k)}-1 \Bigr)P_{l}(\cos \theta) \right]
\nonumber \\
&=& f(k,\cos \theta).
\label{(1.4)}
\end{eqnarray}
In other words, in order to obtain from the residue theorem the countable 
infinity of contributions to the scattering amplitude, it is not enough to
consider the contour integral of a rational function. One needs instead 
the ratio of a polynomial and a cosine function. The latter will also contribute
$(-1)^{l}$ factors that turn $P_{l}(-\cos \theta)$ into
$P_{l}(\cos \theta)$, eventually.
The integral on the left-hand side of Eq. (1.4) can be rotated with
profit to the imaginary axis ${\rm Im}(\lambda)$, going from 
$-i \infty$ to $+i \infty$. The resulting integral is
therefore a Watson transform \cite{DR1965,W1918,S1949,B2018}.

On the other hand, it is well known that central potentials can be studied
with profit in $n$-dimensional Euclidean space ${\bf R}^{n}$ \cite{C1990}. 
In section $2$, relying upon our previous work \cite{E1998}, we consider
the role of the $\lambda$-parameter in the Sch\"{o}dinger equation for
stationary states. Sections $3$ and $4$ are devoted to the cases
real $l$ and complex $n$, complex $l$ and complex $n$, respectively.
Concluding remarks are made in section $5$.

\section{The $\lambda$ parameter}
\setcounter{equation}{0}

As is shown, for example, in Refs. \cite{C1990,E1998}, the Schr\"{o}dinger
equation for stationary states in $n$-dimensional Euclidean space (whose
consideration is inspired by the large-$n$ formalism in quantum mechanics)
leads to the following equation for the radial part $\psi(r)$ of
stationary states:
\begin{equation}
\left[{d^{2}\over dr^{2}}+{(n-1)\over r}{d \over dr}
+{2m \over {\hbar}^{2}}(E-U(r))-{l(l+n-2)\over r^{2}}\right]\psi(r)=0.
\label{(2.1)}
\end{equation}
This equation must be solved with suitable boundary conditions as
$r \rightarrow 0^{+}$ and as $r \rightarrow \infty$, and the full stationary
state reads eventually as
$$
\psi(r)g(\theta_{1},...,\theta_{n-1}),
$$
where $g$ is the generalized spherical harmonic. At this stage, a technique
already developed by Liouville \cite{WW1927} for studying linear differential
equations with variable coefficients suggests setting
\begin{equation}
\psi(r)=r^{\beta}\varphi(r),
\label{(2.2)}
\end{equation}
where $\beta=\beta(n)$ should be chosen in such a way that the resulting
differential equation for the unknown function $\varphi$ has vanishing
coefficient of ${d \varphi \over dr}$. One then finds that the desired
$\beta(n)$ should solve the algebraic equation of first degree
\begin{equation}
2\beta+(n-1)=0 \Longrightarrow \beta=-{(n-1)\over 2}.
\label{(2.3)}
\end{equation}
On defining
\begin{equation}
k^{2} \equiv {2mE \over {\hbar}^{2}}, \;
V(r) \equiv {2m \over {\hbar}^{2}}U(r),
\label{(2.4)}
\end{equation}
Eqs. (2.1)-(2.4) lead to the following second-order equation for $\varphi(r)$:
\begin{equation}
\left[{d^{2}\over dr^{2}}-\left(l(l+n-2)+{1 \over 4}(n^{2}-4n+3)\right)
{1 \over r^{2}}+k^{2}-V(r)\right]\varphi(r)=0.
\label{(2.5)}
\end{equation}
Interestingly, by exploiting the identity \cite{E1998}
\begin{equation}
l(l+n-2)=\left(l+{(n-2)\over 2}\right)^{2}-{1 \over 4}(n-2)^{2},
\label{(2.6)}
\end{equation}
one finds that
\begin{equation}
l(l+n-2)+{1 \over 4}(n^{2}-4n+3)=\left(l+{(n-2)\over 2}\right)^{2}
-{1 \over 4},
\label{(2.7)}
\end{equation}
and hence Eq. (2.5) takes eventually the form
\begin{equation}
\left[{d^{2}\over dr^{2}}+k^{2}-{\left(\lambda^{2}-{1 \over 4}\right)\over r^{2}}
-V(r) \right]\varphi(r)=0,
\label{(2.8)}
\end{equation}
having defined
\begin{equation}
\lambda=\lambda(l,n) \equiv l+{(n-2)\over 2}.
\label{(2.9)}
\end{equation}
With the modern language of quantum mechanics one can say that a unitary
operator $U$ has been built that realizes the map \cite{D2016}
$$
U:L^{2}({\bf R}^{+},r^{n-1}dr) \rightarrow
L^{2}({\bf R}^{+},dr).
$$
The parameter denoted by $\nu(l)$ in Ref. \cite{D2016} coincides with our
$\lambda$ defined in Eq. (2.9), and we refer the reader to Ref. \cite{D2016}
for many important results obtained with arbitrary integer values
of $n$ in Eq. (2.9).

\section{Real $l$ and complex $n$}
\setcounter{equation}{0}

Since the formalism suggests considering complex values of $\lambda$, and
Eq. (2.8) bears full resemblance with the equation for $\varphi(r)$ in
${\bf R}^{3}$ \cite{DR1965}, we can in particular achieve complex values
of $\lambda$ by keeping $l$ real while allowing for complexified values
of $n$. In other words, given a pair $(l_{1},n_{1})$ where
$l_{1} \in {\bf C},n_{1}=3$ (this is the case studied in Refs.
\cite{R1959,BLR1962,DR1965}), and another pair $(l_{2},n_{2})$ where
$l_{2} \in {\bf R},n_{2} \in {\bf C}$, if we require that
\begin{equation}
\lambda(l_{1},n_{1}=3)=\lambda(l_{2},n_{2}),
\label{(3.1)}
\end{equation}
we find
\begin{equation}
{\rm Re}(n_{2})=2{\rm Re}(l_{1})+3-2l_{2},
\label{(3.2)}
\end{equation}
\begin{equation}
{\rm Im}(n_{2})=2{\rm Im}(l_{1}).
\label{(3.3)}
\end{equation}
If such conditions are fulfilled, we can import all results obtained in Ref.
\cite{DR1965}, and hence write that, at large $| \lambda|$ with 
${\rm Re}(\lambda)=l_{2}+{1 \over 2}{\rm Re}(n_{2})-1 >0$, along the ray
${\rm arg}(\lambda)=\sigma$, one finds
\begin{equation}
\left | {\lambda P_{\lambda -{1 \over 2}}(-\cos \theta) \over 
\cos (\pi \lambda)} \right | \leq 
{\kappa_{1} \sqrt{| \lambda |} \over \sqrt{|\sin \theta|}}
{\rm exp} \left \{-|{\rm Re}(\theta){\rm Im}(\lambda)| 
+{\rm Im}(\theta) {\rm Re}(\lambda) \right \}.
\label{(3.4)}
\end{equation}
Moreover, if the potential $V(r)$ is Yukawian\footnote{Recall from
Ref. \cite{DR1965} that the potential $V(z)$ 
is Yukawian if one can express it in the form
$$
V(z)=\int_{m}^{\infty}C(\mu)e^{-\mu z}d\mu,
$$
where $C(\mu)$ is a distribution. Other relevant cases occur when 
$C(\mu)$ is a function of bounded variation, for which the potential
reads as
$$
V(z)=\int_{m}^{\infty}{e^{-\mu z}\over z}dC_{\mu},
$$
or when $C$ is an absolutely continuous function with first derivative
$C'(\mu)=\sigma(\mu)$, for which $V(z)$ takes the form
$$
V(z)=\int_{m}^{\infty}{e^{-\mu z}\over z}\sigma(\mu)d\mu.
$$
The Yukawa potential is recovered from the first formula when
$C(\mu)$ is just a constant.}, then for large enough $|\lambda|$, but
${\rm arg}(\lambda) \not = {\pi \over 2}$, one finds \cite{DR1965}
\begin{equation}
\left | e^{2i \delta(\lambda,k)}-1 \right | \leq \kappa_{2}
{\rm exp}[-\alpha {\rm Re}(\lambda)],
\label{(3.5)}
\end{equation}
where $\alpha$ is such that $\cosh(\alpha)=1+{m^{2}\over 2k^{2}}$.
By virtue of (3.4) and (3.5) one then obtains the important 
majorization \cite{DR1965}
\begin{equation}
\left | 2k Q_{k}(\lambda) \right |
\leq \kappa \sqrt{|\lambda|}
{\rm exp} \left \{ -|{\rm Re}(\theta){\rm Im}(\lambda)|
+{\rm Im}(\theta){\rm Re}(\lambda)-\alpha {\rm Re}(\lambda)
\right \}.
\label{(3.6)}
\end{equation}

\section{Complex $l$ and complex $n$}
\setcounter{equation}{0}

The definition (2.9) suggests investigating the more general case when both
$l$ and $n$ are complex. The resulting framework is non-trivial, because we
are then asking whether Eq. (1.1) can be viewed as arising from the residue
of a double integral of a function of two complex variables, $l$ and $n$. 
In other words, the Regge formslism in the form developed in the sixties and
until now is part of a broader framework, and it is our task to define
properly such a framework and then investigate its potentialities. 

\subsection{Double integrals in ${\bf C}^{2}$}
For this purpose, we begin by describing some basic properties of complex
integration of expressions like
\begin{equation}
I=\int \int_{\sigma}F(\xi,\eta)d\xi \; d\eta,
\label{(4.1)}
\end{equation}
following the pioneering work of Poincar\'e \cite{P1887}, who was the
undisputed founder of this research line (Leray provided the modern rigorous
theory in Ref. \cite{Leray}, for which we refer also to
Ref. \cite{AMS}).

A double integral must be evaluated on a surface, denoted by $\sigma$ in (4.1).
If the function $F$ can be expressed by the ratio 
${p(\xi,\eta)\over q(\xi,\eta)}$, it becomes of course infinite when the
denominator $q(\xi,\eta)$ vanishes. Such a condition leads in turn to the
pair of equations (we set $\xi=x+iy,\eta=z+it$)
\begin{equation}
{\rm Re} \; q(\xi,\eta)=q_{1}(x,y,z,t)=0,
\label{(4.2)}
\end{equation}
\begin{equation}
{\rm Im} \; q(\xi,\eta)=q_{2}(x,y,z,t)=0.
\label{(4.3)}
\end{equation}
Equations (4.2) and (4.3) are the defining equations of the singular surface
$\Sigma_{s}$, and we require that the integration surface $\sigma$ has
empty intersection with $\Sigma_{s}$. 

Let $(\lambda,\mu,\nu)$ be the coordinates of a point in ${\bf R}^{3}$,
and let us consider an algebraic surface $S$, or portion of an algebraic
surface, upon which the point $(\lambda,\mu,\nu)$ lies. On writing
\begin{equation}
x=\varphi_{1}(\lambda,\mu,\nu), \;
y=\varphi_{2}(\lambda,\mu,\nu), \;
z=\varphi_{3}(\lambda,\mu,\nu), \;
t=\varphi_{4}(\lambda,\mu,\nu), 
\label{(4.4)}
\end{equation}
we may assume that $\varphi_{1},\varphi_{2},\varphi_{3},\varphi_{4}$
are rational functions of $\lambda,\mu,\nu$ whose denominator never vanishes. 
By construction, when the point $(\lambda,\mu,\nu)$ describes in ${\bf R}^{3}$
the surface (or portion of surface) $S$, the point $(x,y,z,t)$ describes a surface
(or portion of surface) $S'$. The resulting surface $S'$ is defined by the surface 
$S$ and by the four fundamental functions $\varphi_{1},\varphi_{2},\varphi_{3}$
and $\varphi_{4}$. We can also work under weaker assumptions, and hence assume
that $\varphi_{1},\varphi_{2},\varphi_{3}$ and $\varphi_{4}$ in (4.4) are just
single-valued analytic functions of $\lambda,\mu,\nu$ at least in a certain
subset of ${\bf R}^{3}$, in such a way that the surface $\sigma$ that represents
the integration surface is entirely contained within such a subset.

In order to ascribe a meaning to a formula like (4.1), we must also define the
concepts of inward- or outward-pointing. For this purpose, we can imagine an
observer ${\cal O}$ standing on the surface $S$ with his/her head directed towards
the interior of the surface. The position of such an observer defines the 
sense of integration. Following Poincar\'e \cite{P1887}, the sense of integration
is positive (resp. negative) if the observer's head is directed towards the
exterior (resp. interior). If the surface $S$ is not closed, we can nevertheless
agree that one of the sides is the exterior, while the other side is then said to
represent the interior. On setting $F=P+iQ$, the formal manipulation of (4.1)
would yield
\begin{eqnarray}
\; & \; &
I = \int \int (P+iQ)(dx+i \; dy)(dz+i \;dt)
\nonumber \\
&=& \int \int [(P+iQ)dx \; dz+(iP-Q)dx \; dt 
\nonumber \\
&+& (iP-Q)dy \; dz
-(P+iQ)dy \; dt].
\label{(4.5)}
\end{eqnarray}
If we now assume that two auxiliary variables $u,v$ exist such that, at all
points of the surface $S$, the three coordinates $\lambda,\mu,\nu$ are
holomorphic functions of $u$ and $v$, the desired integral (4.5) becomes
the ordinary double integral
\begin{eqnarray}
\; & \; & \int \int \left[(P+iQ){\partial (x,z) \over \partial (u,v)}
+(iP-Q){\partial (x,t) \over \partial (u,v)} \right .
\nonumber \\
&+& \left . (iP-Q){\partial (y,z) \over \partial (u,v)}
-(P+iQ){\partial (y,t) \over \partial (u,v)} \right]
du \; dv,
\label{(4.6)}
\end{eqnarray}
extended over all values of $u$ and $v$, and which correspond to the different
points of the surface $S$ described by $(\lambda,\mu,\nu)$ as they vary.
More precisely, we may have to decompose the surface $S$ into a number of
surfaces of smaller measure, within each of which two variables $u,v$ exist
such that $\lambda,\mu,\nu$ are holomorphic functions of $u$ and $v$.

Note that the integral (4.6) changes sign when a permutation of $u$ and $v$
is performed. In order to fix a convention, let us imagine that $u,v$ represent 
the coordinates of a point in the plane. Let us also imagine that the point
$(\lambda,\mu,\nu)$ describes on the surface $S$ a very small closed contour
$C$ about the feet of the observer ${\cal O}$, and that this observer sees 
this point as describing anticlockwise the contour $C$. The corresponding point 
$(u,v)$ describes in its plane another closed contour $C'$. The contour
$C'$ must be traveled anticlockwise as it occurs for the contour $C$ on
$S$, assuming that the $u>0$ and $v>0$ axes are arranged as the $x>0$
and $y>0$ axes. The integrability conditions for (4.6) read
\begin{equation}
{\partial \over \partial x}(iP-Q)
-{\partial \over \partial y}(P+iQ)=0,
\label{(4.7)}
\end{equation}
\begin{equation}
-{\partial \over \partial x}(P+iQ)
-{\partial \over \partial y}(iP-Q)=0,
\label{(4.8)}
\end{equation}
\begin{equation}
{\partial \over \partial t}(P+iQ)
-{\partial \over \partial z}(iP-Q)=0,
\label{(4.9)}
\end{equation}
\begin{equation}
{\partial \over \partial t}(iP-Q)
+{\partial \over \partial z}(P+iQ)=0.
\label{(4.10)}
\end{equation}

Let us consider the portions of surfaces $S$ and $S'$ bounded by the same
contour $C$, assuming that both portions lie on the three-space with 
coordinates $\lambda,\mu,\nu$. Let the two integration surfaces be defined by
$S$ and $S'$, respectively, but having the same fundamental equations (4.4).
If the surface $S$ can, by continuous deformation, coincide with $S'$, and
if during such a deformation it never happens that $F=P+iQ$ becomes infinite
or discontinuous at a point of the integration surface, the integral evaluated
on $S$ equals the integral evaluated on $S'$.

On the other hand, the singular surfaces where $|F|=\infty$ or $F$ is discontinuous,
are described by a pair of equations (cf. (4.2) and (4.3))
\begin{equation}
\phi_{1}(x,y,z,t)=0, \;
\phi_{2}(x,y,z,t)=0.
\label{(4.11)}
\end{equation}
If we here re-express $x,y,z,t$ by means of Eqs. (4.4), the two equations for
singular surfaces reduce to the relations
\begin{equation}
f_{1}(\lambda,\mu,\nu)=0, \;
f_{2}(\lambda,\mu,\nu)=0
\label{(4.12)}
\end{equation}
among $\lambda,\mu$ and $\nu$. Such equations describe skew curves in
${\bf R}^{3}$ with coordinates $\lambda,\mu,\nu$, and such curves are
said to be the singular curves, because they are the set of all points 
$(\lambda,\mu,\nu)$ of ${\bf R}^{3}$ where $|F|=\infty$ or $F$ becomes
discontinuous. To sum up, one of the following cases may occur:
\vskip 0.3cm
\noindent
(i) The two portions of surface $S$ and $S'$ being bounded by the same
contour $C$, they divide the $3$-space with coordinates $\lambda,\mu,\nu$
into two regions, one internal and the other external. If in the internal
region there is no point belonging to singular curves, the integral (4.1)
evaluated on $S$ equals the integral evaluated on $S'$.
\vskip 0.3cm
\noindent
(ii) If the surface $S$ is closed, it divides the $3$-space with coordinates
$\lambda,\mu,\nu$ into two regions; if within $S$ there is no point lying
on singular curves, the integral evaluated on $S$ vanishes.
\vskip 0.3cm
\noindent
(iii) If the surface $S'$ is closed as $S$ and entirely contained 
within $S$, and if in the space between $S$ and $S'$ there is no point
lying on singular curves, the integral evaluated on $S$ equals the
integral evaluated on $S'$.
\vskip 0.3cm
\noindent
(iv) If two surfaces $S$ and $S'$, both closed and belonging to $3$-space
with coordinates $\lambda,\mu,\nu$, contain inside them the same singular
curves and the same portions of singular curves, the integral (4.1) 
evaluated on $S$ equals the integral evaluated on $S'$.

In the simplest possible terms, a double integral taken along a closed surface
$S$ can only depend on the singular curves contained within such a 
surface \cite{P1887}. We therefore consider the double integral (4.1) taken
over a closed surface $S$, the observer ${\cal O}$ being outward-pointing.
Inspired by the previous sections, we assume that
\begin{equation}
F(\xi,\eta)={P(\xi,\eta)\over Q(\xi,\eta)},
\label{(4.13)}
\end{equation}
where $P$ is a polynomial while $Q$ has zeros corresponding to the singular
curves, i.e.,
\begin{equation}
Q[\varphi_{1}(\lambda,\mu,\nu)+i \varphi_{2}(\lambda,\mu,\nu),
\varphi_{3}(\lambda,\mu,\nu)+i \varphi_{4}(\lambda,\mu,\nu)]=0.
\label{(4.14)}
\end{equation}
Unlike the theory developed by Poincar\'e\footnote{The assumption of
Poincar\'e is not however too restrictive, since a theorem holds, according 
to which a function that is meromorphic over the whole of ${\bf C}^{2}$,
including the point at infinity, is a rational function \cite{RC1947}.}, 
our $Q(\xi,\eta)$ is not a polynomial, but remains responsible for 
the meromorphic nature of $F(\xi,\eta)$.

Let us assume that the surface $S$ contains two singular curves, e.g., $C$ 
and $C'$. We can always build in ${\bf R}^{3}$ with points $(\lambda,\mu,\nu)$
two closed surfaces $\Sigma$ and $\Sigma'$, both contained within $S$, and such
that $\Sigma$ (resp. $\Sigma'$) contains only $C$ (resp. $C'$). The integral
(4.1), when taken over $S$, is then the sum of the integral taken over
$\Sigma$ and the integral taken over $\Sigma'$, the observer ${\cal O}$ being
always directed outwards. Eventually, we revert to the case where the integration
surface $S$ contains just one singular curve $C$.

It is not necessary for the various surfaces $S$ to belong to the same space
with points $(\lambda,\mu,\nu)$. Let us therefore consider a particular space
with points denoted instead by $(\lambda',\mu',\nu')$ that nevertheless 
contains the singular curve $C$ and, within such a space, a closed surface
$\Sigma$ enclosing such a curve. The idea is to choose this surface in such
a way that integration of $F$ over $S$ equals, up to a sign, the integration
of $F$ over $\Sigma$. 

Indeed, since the singular curve $C$ is closed, its equations can be always
written in the form
\begin{equation}
x=\psi_{1}(\omega), \; y=\psi_{2}(\omega), \;
z=\psi_{3}(\omega), \; t=\psi_{4}(\omega),
\label{(4.15)}
\end{equation}
the $\psi$'s being periodic functions of the parameter $\omega$. Following
Poincar\'e \cite{P1887}, their period is here set to $2 \pi$. As a next
step, we introduce two auxiliary parameters $\rho$ and $\varphi$ for which
\begin{equation}
\lambda'=\cos \omega(1+\rho \cos \varphi), \; 
\mu'=\sin \omega (1+\rho \cos \varphi), \;
\nu'=\rho \sin \varphi,
\label{(4.16)}
\end{equation}
\begin{equation}
x=\psi_{1}(\omega), \; y=\psi_{2}(\omega), \;
z=\psi_{3}(\omega)+\rho \cos \varphi, \;
t=\psi_{4}+\rho \sin \varphi.
\label{(4.17)}
\end{equation}
By virtue of (4.17) and (4.16), $x,y,z,t$ are functions of    
$\omega,\rho,\varphi$ and, eventually, of $\lambda',\mu',\nu'$. At this stage,
however, $x,y,z,t$ are single-valued analytic functions of $\omega,\rho,\varphi$ 
but not of $\lambda',\mu',\nu'$. However, if we agree that 
$\rho \in ]0,1[$, then to a set of values $\lambda',\mu',\nu'$ there corresponds
a unique set of values of $\rho,\cos \omega,\sin \omega,\cos \varphi,
\sin \varphi$, and hence a unique set of values of $x,y,z,t$. By virtue of
this restriction, $x,y,z,t$ become eventually single-valued analytic functions 
of $\lambda',\mu',\nu'$. To a point $(\lambda',\mu',\nu')$ satisfying the condition 
$\rho <1$ and hence lying within a certain torus, there corresponds one and
only one point $(x,y,z,t)$. 

Within this representation scheme, the curve $C$ is represented by the
circle ($\rho=0$)
\begin{equation}
{\lambda'}^{2}+{\mu'}^{2}=1, \; \nu'=0.
\label{(4.18)}
\end{equation}
The Poincar\'e choice for the surface $\Sigma$ is the torus
\begin{equation}
\rho=\rho_{0}, \; \rho_{0} \in ]0,1[
\label{(4.19)}
\end{equation}
that wraps the curve $C$. We can now decompose the integration over $\Sigma$ into
two steps. First, we integrate over $\eta$, regarding $\xi$ as an arbitrary 
parameter. Hence we have
\begin{equation}
z=\psi_{3}(\omega)+\rho_{0}\cos \varphi, \;
t=\psi_{4}(\omega)+\rho_{0}\sin \varphi.
\label{(4.20)}
\end{equation}
If $\xi$ is viewed for a moment as a constant, $\omega$ is then also a constant;
the same holds for $\rho_{0}$, $\varphi$ being the only independent variable,
and we have
\begin{equation}
\eta=\psi_{3}+i \psi_{4}+\rho_{0}e^{i \varphi},
\label{(4.21)}
\end{equation}
which shows that the point $\eta$ describes in the $\eta$-plane a circle of
radius $\rho_{0}$ centred at the point $\psi_{3}+i \psi_{4}$. The integral
\begin{equation}
I=\int F(\xi,\eta)d\eta,
\label{(4.22)}
\end{equation}
is therefore equal to $2i \pi$ times the residue of the function $F(\xi,\eta)$
(viewed as a function of $\eta$ only) with respect to the point 
$\psi_{3}+i \psi_{4}$.

Along the singular curve $C$, the denominator $Q(\xi,\eta)$ in (4.13) vanishes:
\begin{equation}
Q(\xi,\eta)=Q(\psi_{1}+i \psi_{2},\psi_{3}+i \psi_{4})=0.
\label{(4.23)}
\end{equation}
In the final stage, one has to integrate with respect to $\xi$ by letting $\omega$
vary from $0$ to $2\pi$, i.e., upon following the whole singular curve $C$. We are
therefore led to studying the integral
\begin{equation}
J=\int 2i\pi {\rm Res} \left(
{P(\xi,\eta)\over Q(\xi,\eta)}\right)d\xi,
\label{(4.24)}
\end{equation}
along the curve $C$, $\eta$ being related to $\xi$ by the equation (4.23).

\subsection{The case of the scattering amplitude}

Since Poincar\'e restricted himself to rational functions $F={P \over Q}$, his
condition $Q=0$ was an algebraic equation, unlike our case, where we propose
to look for functions $P$ and $Q$ such that, by virtue of Eq. (1.2),
\begin{equation}
2i \pi {\rm Res} {P(l,n(l))\over Q(l,n(l))}
={1 \over 2k} \left(e^{2i \delta(\lambda,k)}-1 \right)
{\lambda P_{\lambda -{1 \over 2}}(-\cos \theta) \over 
\cos (\pi \lambda)},
\label{(4.25)}
\end{equation}
where on the right-hand side $\lambda$ is defined as in (2.9) but is viewed as a
single complex variable.

\section{Concluding remarks}
\setcounter{equation}{0}

The complex angular momentum formalism is still finding applications, not only
to particle physics \cite{DM2018}, but also to the scattering of scalar, electromagnetic
and gravitational waves by a Schwarzschild black hole \cite{FHa,FHb}. 
From the point of view of general formalism, Regge's method was remarkable because
his use of Watson transform made it possible to elucidate analytic properties of the
scattering amplitude. On the other hand, from the point of view of particle
physics, if the variable $l$ is interpreted as the intrinsic angular momentum, 
a Regge singularity establishes a relation between spin and mass of the exchanged
particle \cite{B2018}. In modern times, this picture is possibly superseded by 
the gauge theories of fundamental interactions, but one may instead appreciate the
interest of our Sect. $3$, where complex $\lambda$ is achieved by complexifying
$n$ only. The complexification of $n$ may seem counterintuitive, but is perfectly
consistent with the residue calculation in Eq. (1.3), and in our opinion it would
be a lost opportunity to regard it as counterintuitive or a pure artifice devoid of 
physical interest.

In our paper we have then undertaken the task of preparing the ground for
a derivation of the non-relativistic scattering amplitude from double integrals 
in the complex domain. Our contribution is so far at the level of ideas, but
the picture outlined in this manuscript adds evidence in favour of
new applications of complex analysis to quantum physics being in sight.
 
\section*{acknowledgements}
The author is gra\-te\-ful to the Di\-par\-ti\-men\-to di Fi\-si\-ca 
``Et\-to\-re Pan\-ci\-ni'' for hos\-pi\-ta\-li\-ty a\-nd sup\-po\-rt,
and to Roland Rosenfelder and Gabriele Gionti for correspondence.

\end{document}